\documentclass[twocolumn,showpacs,aps,prl,superscriptaddress]{revtex4}

\usepackage{graphicx}
\usepackage{dcolumn}
\usepackage{amsmath}
\usepackage{epsfig}
\usepackage{verbatim}

\input babarsym

\newcommand{\BABARPubYear}    {07}
\newcommand{\BABARPubNumber}  {049}

\newcommand{\SLACPubNumber} {12728}

\def\figurebox#1#2#3{%
    \def\arg{#3}%
    \ifx\arg\empty
    {\hfill\vbox{\hsize#2\hrule\hbox to #2{\vrule\hfill\vbox to #1{\hsize#2\vfill}\vrule}\hrule}\hfill}%
    \else
    {\hfill\epsfbox{#3}\hfill}%
    \fi}

\def\babar{\mbox{\slshape B\kern-0.1em{\smaller A}\kern-0.1em
    B\kern-0.1em{\smaller A\kern-0.2em R}}}
\def\DDK {\ensuremath{\Dbar^{(*)} D^{(*)}  K}}

\def\DsOneRes {\ensuremath{D^+_{s1}(2536)}}

\def\PsiRes {\ensuremath{\psi(3770)}}
\def\XRes {\ensuremath{X(3872)}}

\def\to {\rightarrow}

\def\modexi{\ensuremath{\Bu\to \Dzb \Dz \Kp}}

\def\modexiii{\ensuremath{\Bu\to \Dm \Dp \Kp}}

\def\modeivDsOne{\ensuremath{\Bz\to \Dm \DsOneRes}}
\def\modeviiiDsOne{\ensuremath{\Bz\to \Dstarm \DsOneRes}}
\def\modexviiDsOne{\ensuremath{\Bu\to \Dzb \DsOneRes}}
\def\modexxDsOne{\ensuremath{\Bu\to \Dstarzb \DsOneRes}}
\def\modeviiDsOne{\ensuremath{\Bz\to \Dm \DsOneRes}}
\def\modexDsOne{\ensuremath{\Bz\to \Dstarm  \DsOneRes}}
\def\modexviiiDsOne{\ensuremath{\Bu\to \Dzb  \DsOneRes}}
\def\modexxiDsOne{\ensuremath{\Bu\to \Dstarzb  \DsOneRes}}
\def\modexiPsi{\ensuremath{\Bu\to \PsiRes \Kp}}
\def\modeiiPsi{\ensuremath{\Bz\to \PsiRes K^0}}
\def\modexiiiPsi{\ensuremath{\Bu\to \PsiRes \Kp}}
\def\modeiiiPsi{\ensuremath{\Bz\to \PsiRes K^0}}
\def\modeviX{\ensuremath{\Bz\to \XRes K^0}}
\def\modexxxiX{\ensuremath{\Bu\to \XRes \Kp}}

\def\mes   {\ensuremath{m_{ES}}}
\def\de   {\ensuremath{\Delta E}}

\begin{document}

 \preprint{\babar-PUB-\BABARPubYear/\BABARPubNumber}
 \preprint{SLAC-PUB-\SLACPubNumber}

\begin{flushleft}
\babar-PUB-\BABARPubYear/\BABARPubNumber\\
SLAC-PUB-\SLACPubNumber\\
[10mm]
\end{flushleft}

\title{
{\large \bf
Study of Resonances in Exclusive $B$ Decays to \DDK}
}

%
\author{B.~Aubert}
\author{M.~Bona}
\author{D.~Boutigny}
\author{Y.~Karyotakis}
\author{J.~P.~Lees}
\author{V.~Poireau}
\author{X.~Prudent}
\author{V.~Tisserand}
\author{A.~Zghiche}
\affiliation{Laboratoire de Physique des Particules, IN2P3/CNRS et Universit\'e de Savoie, F-74941 Annecy-Le-Vieux, France }
\author{J.~Garra~Tico}
\author{E.~Grauges}
\affiliation{Universitat de Barcelona, Facultat de Fisica, Departament ECM, E-08028 Barcelona, Spain }
% ...
\collaboration{The \babar\ Collaboration}
\noaffiliation

\date{\today}

\begin{abstract}
We present a study of resonances in exclusive decays of $B$ mesons to \DDK. We report the
observation of the decays $B \to \Dbar^{(*)} \DsOneRes$ where the \DsOneRes\ is
reconstructed in the $D^{*0}K^+$ and $D^{*+}\KS$ decay channels. We report also the
observation of the decays $B \to \PsiRes K$ where the \PsiRes\ decays to $\Dbar^0 D^0$
and $D^-D^+$. In addition, we present the observation of an enhancement for the
$\Dbar^{*0}D^0$ invariant mass in the decays $B \to \Dbar^{*0}D^0K$, at a mass of
$(3875.1 {}^{+0.7}_{-0.5} \pm 0.5) \mevcc$ with a width of $(3.0 {}^{+1.9}_{-1.4} \pm
0.9) \mev$ (the first errors are statistical and the second are systematic). Branching
fractions and spin studies are shown for the three resonances. The results are based on
347~\invfb of data collected with the \babar\ detector at the PEP-II \B factory.
\end{abstract}


\maketitle


In this article, we study the production of \DsOneRes, \PsiRes\ and \XRes\ resonances in
decays of charged and neutral \B mesons to \DDK. Here, $D^{(*)}$ is either a \Dz,
\Dstarz, \Dp or \Dstarp, $\Dbar^{(*)}$ is the charge conjugate of $D^{(*)}$ and $K$ is
either a \Kp or a \KS. Both $\Dbar^{(*)}$ and $D^{(*)}$ are fully reconstructed. Charge
conjugate reactions are assumed throughout this article.

The \DsOneRes\ resonance is the narrow P-wave \Ds meson with $J^P=1^+$ assignment
strongly favored. It can be produced in $B\to \Dbar^{(*)}\DsOneRes$ decays and should
decay dominantly to $\Dstarz \Kp$ and $\Dstarp \Kz$~\cite{pdg}. Evidence for \DsOneRes\
production in $B$ decays was found by \babar\ in the sum of all $\Dbar^{(*)}\Dstarz \Kp$
final states~\cite{patrick} and more recently in the decay $\Bz \to \Dstarm \Dstarp
\KS$~\cite{chunhui}. For most of the $B\to \Dbar^{(*)}\DsOneRes$ modes, only limits have
been placed on the branching fractions~\cite{patrick}. We report herein branching
fraction measurements of $B\to \Dbar^{(*)}\DsOneRes$ decays, through a comprehensive
study of both $\Dbar^{(*)}\Dstarz \Kp$ and $\Dbar^{(*)}\Dstarp \KS$ final states.

The \PsiRes\ meson is a charmonium state with $J^P=1^-$, with a mass just above the open
charm threshold. This meson is thought to be an admixture of the $D$-wave and $S$-wave of
the angular momentum eigenstates of $c\bar{c}$ system. Study of this state in $B$ decays
and branching fraction measurements could provide more information on the structure of
the \PsiRes\ wave function. This resonance decays dominantly to $\Dbar^0 D^0$ and
$D^-D^+$~\cite{pdg}, and was observed in $B$ meson decays by the Belle
experiment~\cite{psiBelle}. We present a study of the $\Dbar D$ mass distribution in
$\Dbar D K$ events.

The \XRes\ resonance was discovered by Belle in the invariant mass distribution of
$J/\psi \pi^+ \pi^-$ produced in $B \to J/\psi \pi^+ \pi^- K$ decays, and was thereafter
confirmed by \babar, D0 and CDF~\cite{X3872Discovery}. This new meson has a mass of
$3871.2 \pm 0.5 \mevcc$ and a natural width less than 2.3~\mev at 90\% confidence level
(C.L.). At present, the quantum numbers compatible with the observations are $J^{PC} =
1^{++}$ or $2^{-+}$~\cite{XCDF}. Recently, Belle showed an excess of events in the
$\Dbar^0 D^0 \pi^0$ invariant mass in the channel $B \to \Dbar^0 D^0 \pi^0 K$, with a
mass of $3875.2 \pm 0.7 {}^{+0.3}_{-1.6} \pm 0.8 \mevcc$~\cite{belleXDDK} (where the
third error is due to the uncertainty in the neutral $D$ mass). The \XRes\ is probably
not a charmonium state, given its measured mass and width, and several alternative
interpretations have been proposed: $\Dbar^{*0} D^0$ molecule, tetraquark state, hybrid
or gluonium states~\cite{X3872Model}. We present a search for \XRes\ decays to
$\Dbar^{*0} D^{0}$.

The measurements reported here use 347~\invfb of data, corresponding to $(383 \pm 4)
\times 10^6$ \BB pairs, collected at the $\Upsilon(4S)$ resonance with the \babar\
detector at the PEP-II \B factory. The \babar\ detector is described in detail
elsewhere~\cite{babar}. We use a Monte Carlo (MC) simulation based on
GEANT4~\cite{geant4} to study the relevant backgrounds and estimate the selection
efficiencies.


The \Bz and \Bu mesons are reconstructed in a sample of hadronic events in the 22
possible \DDK\ modes. The selection criteria are optimized for each final state by
maximizing the significance $S/\sqrt{S+B}$, where $S$ and $B$ refer to the expected
number of signal and background events, based on MC simulation.
The \KS candidates are reconstructed from two oppositely charged tracks consistent with
coming from a common vertex and having an invariant mass within $\pm 9.5\mevcc$ of the
nominal \KS mass. For some channels, depending on the background level, we apply a
requirement on the displacement of the \KS vertex in the plane transverse to the beam
axis of at least 0.2\cm. The \piz candidates are reconstructed from pairs of photons with
energies $E_\gamma>30\mev$ in the laboratory frame that have an invariant mass of
$115<m_{\gaga}<150\mevcc$.
We reconstruct $D$ mesons in the modes $\Dz\to \Km\pip$, $\Km\pip\piz$,
$\Km\pip\pim\pip$, and $\Dp\to \Km\pip\pip$. The $K$ and $\pi$ tracks are required to
originate from a common vertex. Charged kaon identification, based on the Cherenkov angle
and the \dedx\ measurements, is used for $K^+$ from $B$ decays and from most $D$ decays.
The invariant masses of the $D$ candidates are required to be within $\pm 2\sigma$ of the
measured $D$ mass. The measured $D$ mass resolution, $\sigma_D$, is $13\mevcc$ for $D^0
\to K^- \pi^+ \pi^0$ and varies from $5.5$ to $7\mevcc$ for the other modes. The $D^*$
candidates are reconstructed in the decay modes $D^{*+}\to \Dz\pip$, $D^{*+}\to D^+\piz$,
$\Dstarz\to \Dz\piz$, and $\Dstarz\to \Dz\g$. The \piz and the \pip must have momentum
below 450\mevc in the \FourS rest frame, while the \g energy in the laboratory frame must
be greater than 100\mev. The mass difference between the \Dstar and $D$ candidates is
required to be within $3 \mevcc$ of the nominal value~\cite{pdg} for $D^{*+}$ decays ($4
\mevcc$ and $10 \mevcc$ for \Dstarz $\to$ \Dz \piz and \Dstarz $\to$ \Dz $\gamma$,
respectively). The mode $D^{*+}\to D^+\piz$ is used only in the reconstruction of decays
$\Bz \to D^{*-}D^{*+}\KS$ and $\Bu \to D^{*-}D^{*+}\Kp$.

$B$ candidates are reconstructed by combining a $\Dbar^{(*)}$, a $D^{(*)}$ and a $K$
candidate. For most of the modes involving two \Dz mesons, at least one of them is
required to decay to $\Km \pip$. During the optimization process, we remove $D$ decay
modes if the significance for a particular $B$ decay mode improves. In practice, only
$\Dbar^{*0} \Dstarp \KS$ decays benefit from this optimization, for which we remove the
combination with the first $D^0$ meson of the decay chain going to $\Km\pip$ and the
second going to $\Km\pip\pim\pip$ (and vice-versa). A mass-constrained kinematic fit is
applied to the intermediate particles (\Dstarz, $D^{*+}$, \Dz, \Dp, \KS, \piz) to improve
their momentum resolution. For the $D^0$ and the $D^*-D^0$ mass differences, we use the
recent CLEO measurement of the $D^0$ mass~\cite{cleoD0}. To suppress the $\epem\to\qqbar$
($q=u,d,s$ and $c$) continuum background, we perform a selection based on the ratio of
the second to zeroth Fox-Wolfram moments of the event~\cite{fox} and on the cosine of the
angle between the thrust axis of the candidate decay and the thrust axis of the rest of
the event. Signal events have $\mes = \sqrt{s/4-p_B^{*2}}$, where $p_B^*$ is the
center-of-mass momentum of the $B$ candidate, compatible with the known $B$ meson mass,
and a difference between the candidate energy and the beam energy in the center-of-mass,
\de, compatible with 0. On average we have about 1.8 signal $B$ candidates per event. If
more than one candidate is selected in an event, we retain the one with the smallest
$|\de|$. In the final selection, we require $|\de|$ to be less than $n\sigma_{\Delta E}$,
where the resolution $\sigma_{\Delta E}$ varies between 7 and 14\mev\ and $n$ is
determined for each mode by the optimization procedure ($n$ ranges from 1 to 4). For each
mode we define a $B$ signal region $\mes^{\mathrm{min}}<\mes<\mes^{\mathrm{max}}$, where
$\mes^{\mathrm{min}}$ ranges from 5.268 to 5.277 \gevcc, and $\mes^{\mathrm{max}}$ from
5.284 to 5.290\gevcc, and a control region, $5.20<\mes<5.26\gevcc$. The signal purity
obtained in the signal region ranges from 17 to 77\%, depending on the mode.


We consider several sources of systematic errors. From the difference between data and
Monte Carlo efficiencies we derive systematic errors of 0.5\% per charged track, 2.2\%
per soft pion from $D^*$ decays, 2.5\% per \KS, 2\% per $K^+$, 3\% per $\pi^0$ and 2\%
per single photon. As an example, the particle identification efficiency for $K^+$ is
measured using a $D^{*+} \to D^0 \pi^+$ data control sample with $D^0 \to K^- \pi^+$.
Other sources of systematic error are also taken into account: limited MC statistics
(1-3\%), the estimate of the total number of $B$ mesons in the data sample (1.1\%),
uncertainties on the $D^{(*)}$ and \KS\ branching fractions (3-8\%) and uncertainties on
the $D^{(*)}$ and $K$ masses (0.5-6\%). In addition, there are uncertainties in the fit
procedure for the different resonances: when fixing a parameter, we repeat the fit
varying the parameter by $\pm1\sigma$ of its error (where $\sigma$ is the 68\% C.L.
standard deviation); these uncertainties include also variation of the background
parametrization. Using the \mes\ control region, we check that the combinatorial
background events do not contain any significant \DsOneRes, \PsiRes\ or \XRes\ signals or
additional peaking structures. Furthermore, for the mass and width measurements, we
include effects from the energy loss in the tracking system, from the uncertainties on
the magnetic field and from the calibration and background of the calorimeter.


\begin{figure*}[htb]
\begin{center}
 \epsfig{file=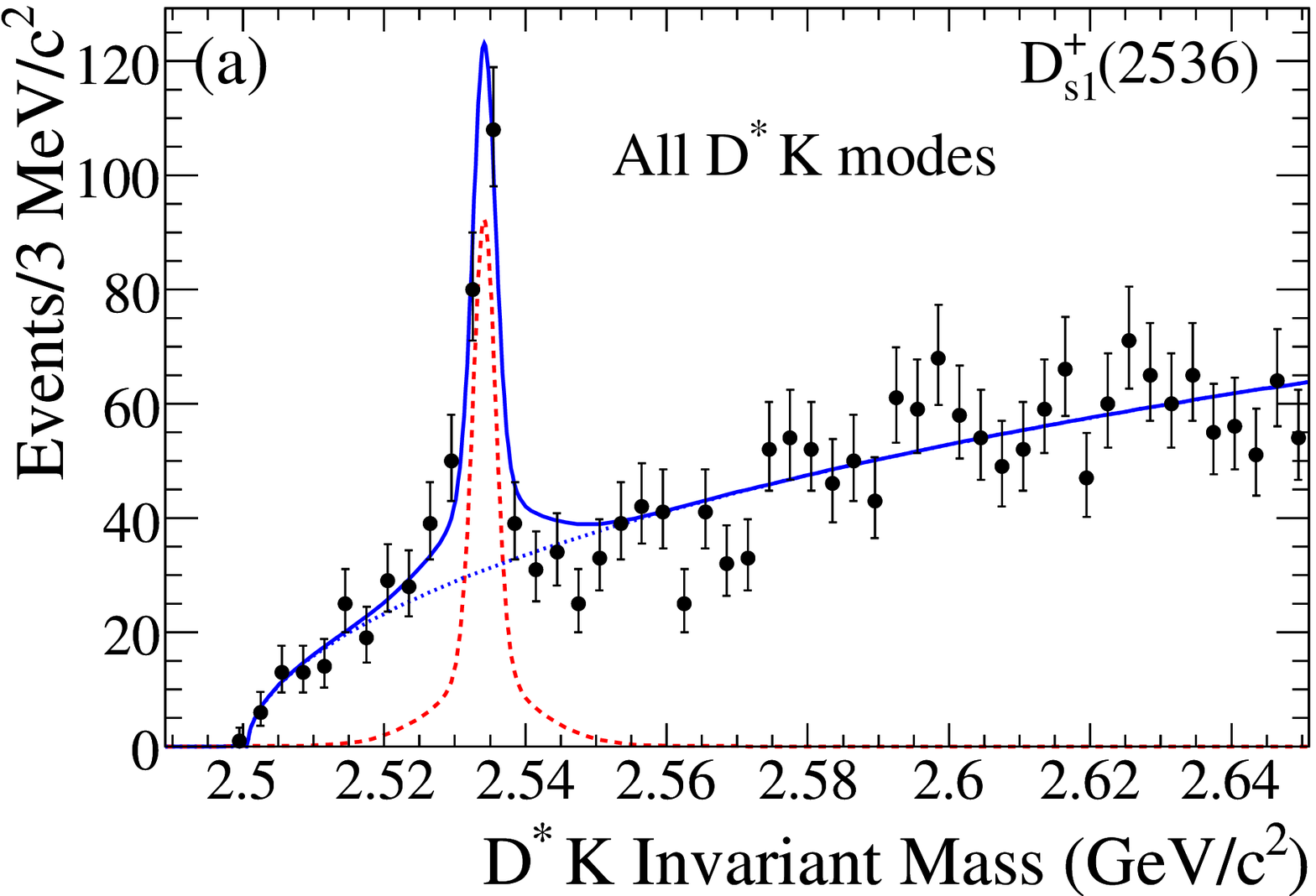,width=5.9cm}
 \epsfig{file=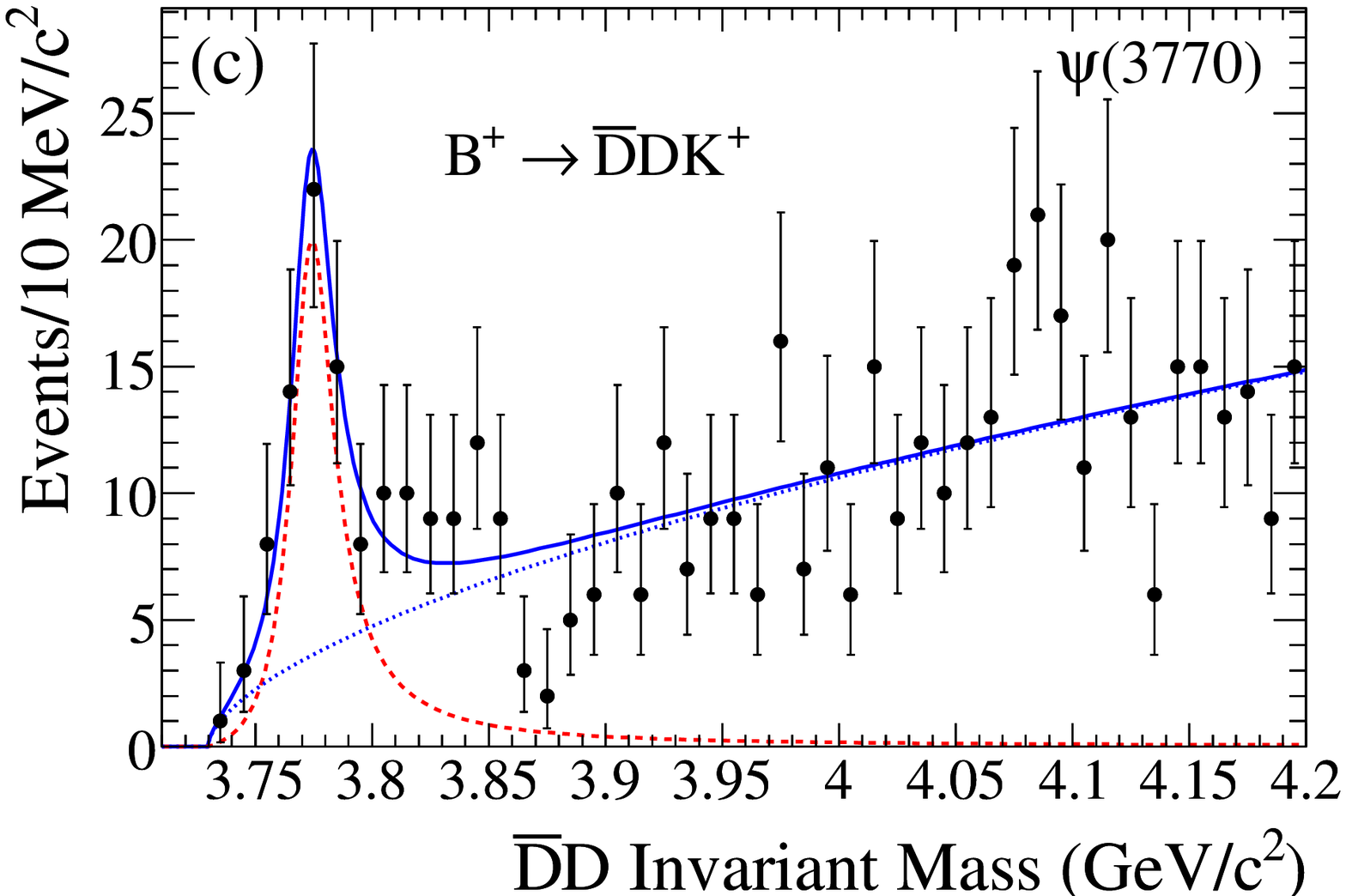,width=5.9cm}
 \epsfig{file=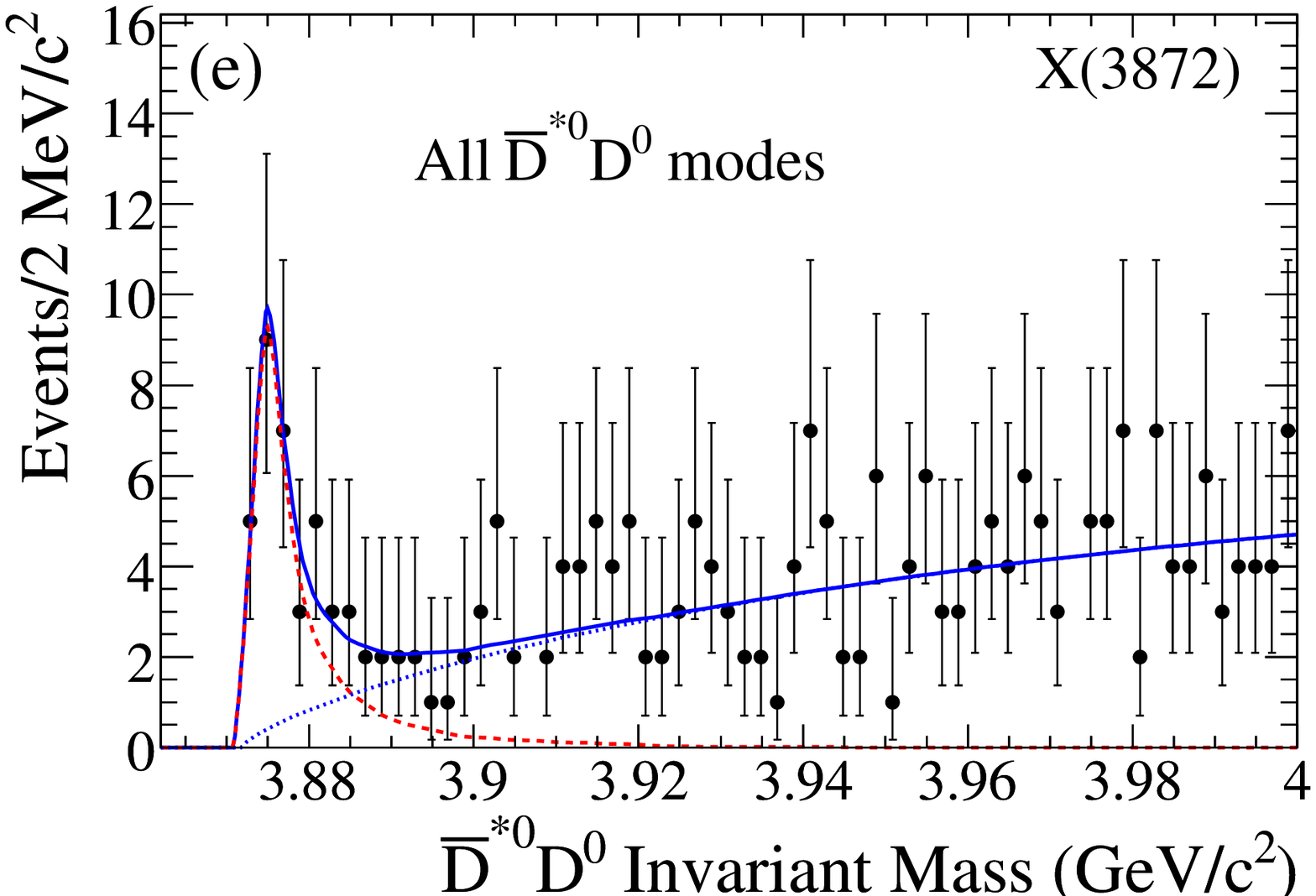,width=5.9cm}
 \epsfig{file=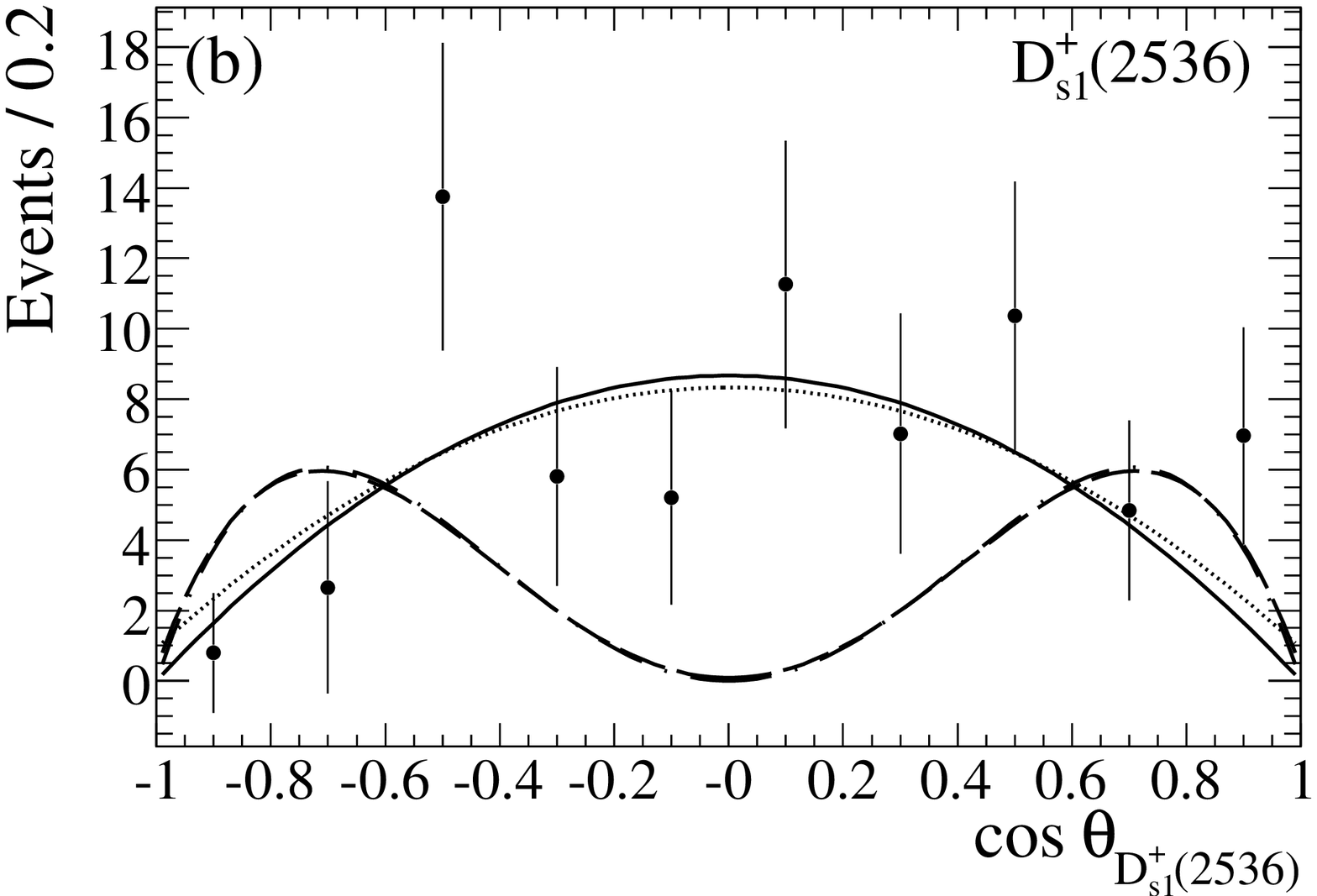,width=5.9cm}
 \epsfig{file=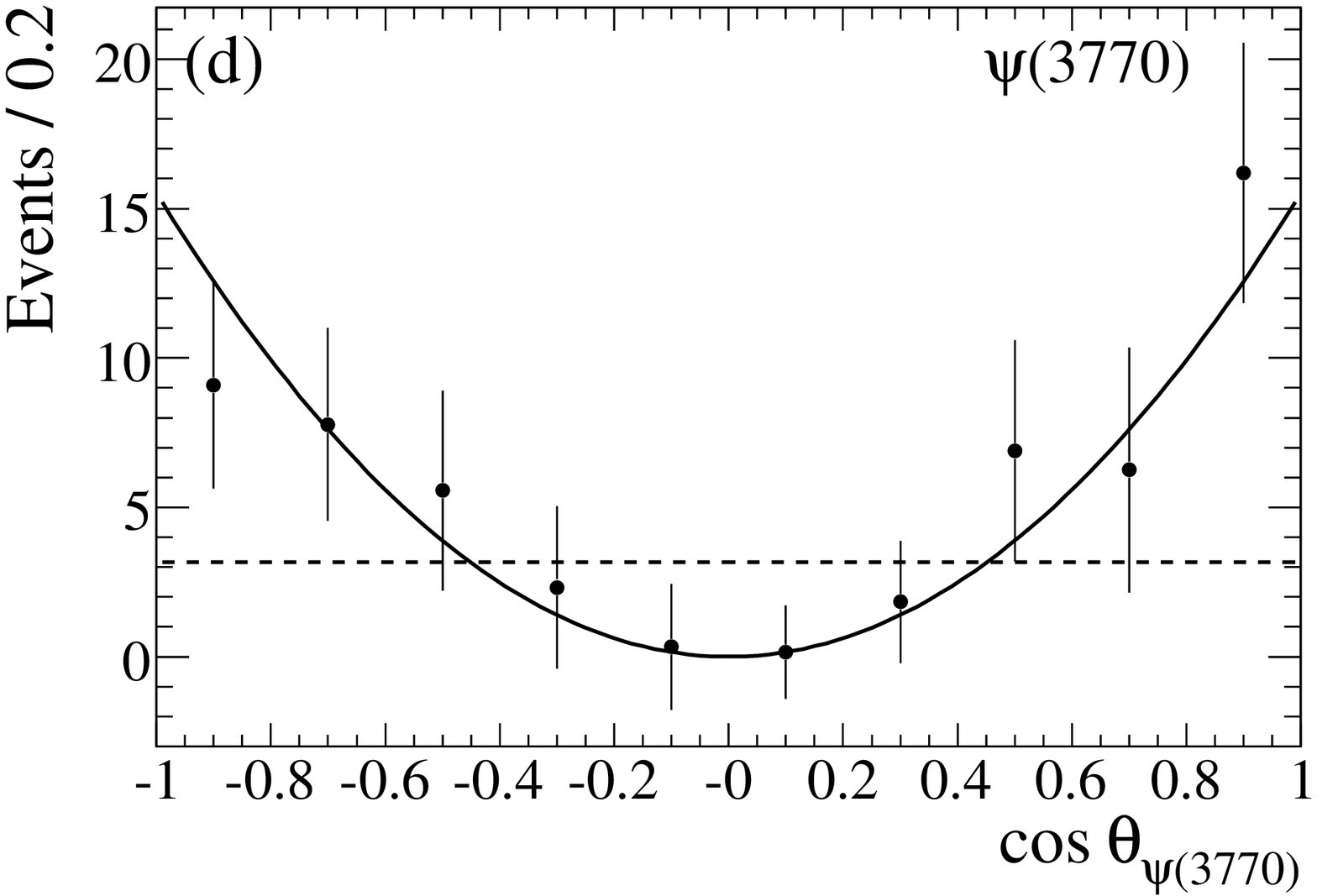,width=5.9cm}
 \epsfig{file=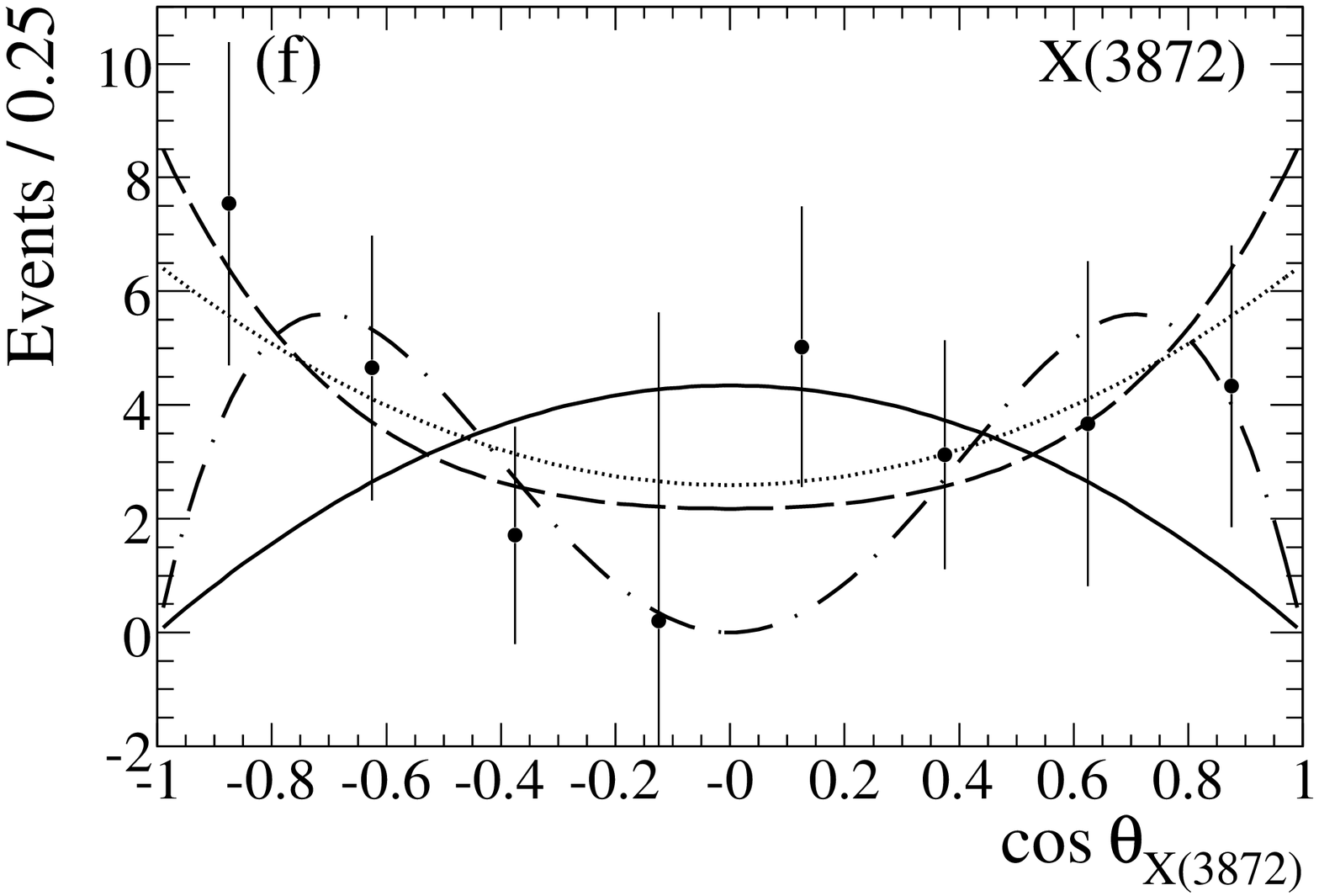,width=5.9cm}

\caption{Top: Invariant mass distributions of $D^{*} K$ (a), $\Dbar D$ (c) and
$\Dbar^{*0} D^0$ (e) in the data for events in the (\mes, \de) $B$ signal region. Points
with statistical errors are data events, the solid line represents the fit to the data,
the dashed line shows the contribution of the \DsOneRes\ (a), \PsiRes\ (c) and \XRes\ (e)
signals, and the dotted line shows the background contribution. Bottom:
background-subtracted and efficiency-corrected helicity angle distributions for the
\DsOneRes\ (b) (containing only the $B$ $\to$ $\Dbar \DsOneRes$ modes), \PsiRes\ (d) and
\XRes\ (f) signals in the data (points with statistical errors) and fitted distributions
for different spin hypotheses:  $J^P=0^+$ (dashed lines), $1^-$ (solid), $1^+$ with
$S$-$D$ wave admixture (dotted), $2^-$ (long-dashed) and $2^+$ (dotted-dashed).}
 \label{figureAll}
\end{center}
\end{figure*}

We search for \DsOneRes\ decaying to $\Dstarp \KS$ and $\Dstarz \Kp$ using $\Bz \to
\Dbar^{(*)-} \Dstarz \Kp$, $\Dbar^{(*)-} \Dstarp \KS$ and $\Bp \to \Dbar^{(*)0} \Dstarz
\Kp$, $\Dbar^{(*)0} \Dstarp \KS$ candidates. We show in Fig.~\ref{figureAll}a the
$\Dstarp \KS$ and $\Dstarz \Kp$ invariant mass distribution for the sum of the eight $B$
modes. The overlaid curve is the result of a unbinned extended maximum likelihood fit.
Since the \DsOneRes\ meson is narrow ($\Gamma_{\DsOneRes} < 2.3\mev$ at 90\%
C.L.~\cite{pdg}), the invariant mass peak is described by the convolution of a
non-relativistic Breit-Wigner with a Gaussian function (called a Voigtian function) and
the background by a threshold function $a (m - m_0)^b \times e^{c (m - m_0)}$, where $m$
is the $D^*K$ invariant mass, $m_0$ is the sum of the $D^*$ and $K$ meson masses, and
$a$, $b$ and $c$ are parameters of the fit. To take into account the large tails in the
reconstruction of $\DsOneRes \to D^{*0} K^+$ decays, the probability density function
(PDF) for the $D^{*0} K^+$ modes is constructed from a sum of a Voigtian function and a
Gaussian function, with a common mean. The \DsOneRes\ mass and yield are floating
parameters in the fit. The natural width of the \DsOneRes\ is fixed to $1\mev$ and varied
from 0.1 to 2.0\mev to estimate systematic errors. The other parameters (the second
Gaussian function and the relative contribution of the Voigtian and the Gaussian
function) are taken from the simulation. A significant signal is observed in each of the
modes separately (see Table~\ref{BRAll}). A fit to the eight $B$ modes gives $182 \pm 19$
events in the peak with a statistical significance of $11.8 \sigma$. We compute an
estimate of the statistical significance calculating \texttt{PROB}$(2(L_0 -
L_{\mathrm{signal}}), N_{\mathrm{dof}})$, where $L_{0(\mathrm{signal})}$ is the minimum
of the likelihood without (with) the signal contribution, $N_{\mathrm{dof}}$ is the
number of free parameters in the signal PDF and \texttt{PROB} is the upper tail
probability of a chi-squared distribution, converting this probability into a number of
standard deviations.

From the \DsOneRes\ yields, we compute cross-feed-corrected branching fractions, using
the signal efficiency and the relative contributions from cross-feed between the
different $B \to \Dbar^{(*)}\DsOneRes$ channels, as obtained from simulated events. The
resulting branching fractions, the efficiencies, including the intermediate branching
fractions, and the internal cross-feed contributions are given in Table~\ref{BRAll}. In
addition to the effects previously mentioned, systematics in the table include
uncertainties from the cross-feed events (0-3\%), underestimate of the MC resolution
(1-10\%) and uncertainty on the \DsOneRes\ natural width (5-18\%). Using only modes
containing $D^{*+}\KS$ in order to minimize the systematic error, we also fit the
\DsOneRes\ mass, $M(\DsOneRes)=(2534.78 \pm 0.31 \pm 0.40) \mevcc$. Our measurement is in
good agreement with the world average~\cite{pdg}.

The helicity angle distribution for the sum of the four $B$ $\to$ $\Dbar \DsOneRes$
modes, determined by fitting the $D^*K$ mass distribution for the \DsOneRes\ yield in ten
$\cos \theta_{\DsOneRes}$ bins, is shown in Fig.\ref{figureAll}b. Here
$\theta_{\DsOneRes}$ is defined as the angle between the $D^*$ direction and the $B$
direction in the \DsOneRes\ frame. We fit the helicity distribution to different
spin-parity hypotheses (one free parameter for $J^P=1^-$ and $2^+$, and two free
parameters for $J^P=1^+$ and $2^-$). The fits to $J^P=1^+$ in pure $S$ wave (flat
distribution, not shown in the figure), $J^P$$=$$1^+$ with $S$-$D$ wave admixture, and
$J^P$$=$$1^-$ are all in good agreement with data, with $\chi^2/$n.d.f. of 15.9/9, 9.3/8
and 9.6/9 respectively. Fits to $J^P$$=$$2^+$ and $2^-$ are disfavored, with
$\chi^2/$n.d.f. of 26.0/9 and 26.0/8 respectively.


We search for $\Bp(\Bz)$ $\to$ $\PsiRes \Kp(\KS)$ with $\PsiRes$ $\to$ $\Dm\Dp$,
$\Dzb\Dz$. In Fig.~\ref{figureAll}c we show the $\Dbar D$ invariant mass for the sum of
\modexi\ and \modexiii\ candidates. No significant signal is observed in the $\Dbar D
\KS$ modes. The curve shown is the result of a fit where the \PsiRes\ peak is described
by a P-wave Breit-Wigner with the mass as a free parameter in the fit and a natural width
fixed to $23\mev$~\cite{pdg}, while the background is described by the threshold function
described previously. We observe $57 \pm 11$ events, with a statistical significance of
$6.4 \sigma$, from which we obtain $M(\PsiRes)=(3775.5 \pm 2.4 \pm 0.5) \mevcc$, in good
agreement with the recent high precision measurement of the BES collaboration~\cite{bes}.
We obtain the branching fractions, or limits, reported in Table~\ref{BRAll}, by fitting
separately the \modexi, $\Bz \to \Dzb \Dz \KS$, \modexiii\ and $\Bz \to \Dm \Dp \KS$
channels, with $M(\PsiRes)$ fixed to the result of the combined fit.

For the two modes with significant signal, we study the \PsiRes\ helicity angle,
$\theta_{\PsiRes}$ (Fig.~\ref{figureAll}d), defined as the angle between the $D^{0/+}$
direction and the $B$ direction in the \PsiRes\ frame. We confirm the spin 1 assignment
of the \PsiRes\ ($\chi^2$/n.d.f.$=$2.9/9). A spin 0 hypothesis gives
$\chi^2$/n.d.f.$=$22.0/9.

\begin{figure}[h]
\begin{center}
\epsfig{file=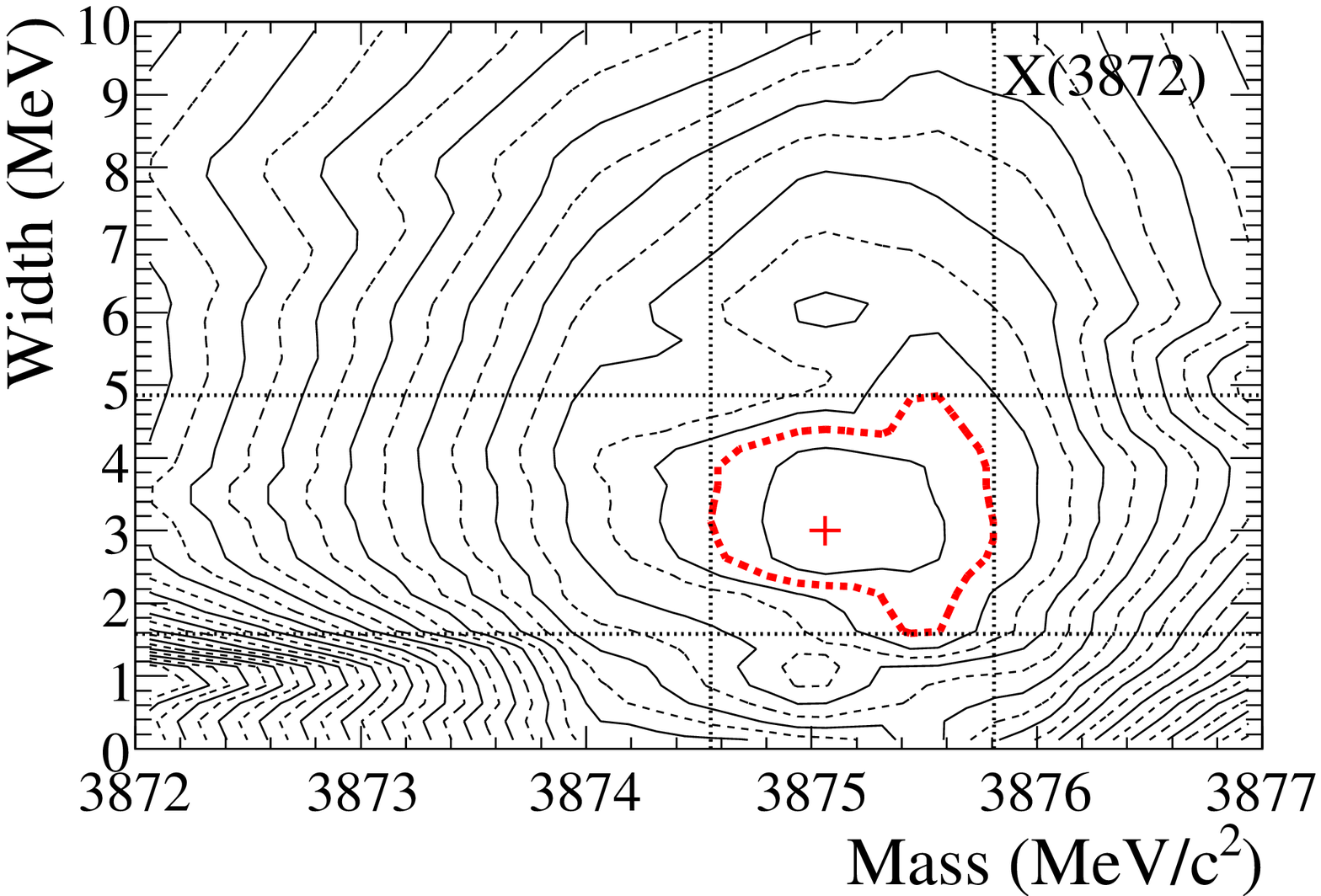,width=5.9cm}
 \caption{Equidistant contours of $\chi^2-\chi^2_{\mathrm{min}}$ versus mass and
width of the simulated \XRes\ signals. The contours are separated by $\chi^2$ values of
0.5. Solid contours represent values starting from 0.5, by step of 1, while dashed
contours represent values starting from 1, by step of 1. The contour at
$\chi^2-\chi^2_{\mathrm{min}} = 1$ is indicated by a wider line. The cross shows the
central values of the \XRes\ mass and width. The vertical and horizontal dotted lines
indicate the errors on the mass and on the width respectively.}
 \label{chi2}
\end{center}
\end{figure}

We search for decays $B$ $\to$ $\X(3872)K$, $X(3872)$ $\to$ $\Dbar^{*0} D^0$ in the
$\Bp(\Bz)$ $\to$ $\Dbar^{0} D^{*0}\Kp (\KS)$ $+$ $\Dbar^{*0} D^{0}\Kp (\KS)$ samples. We
plot the $\Dbar^{*0} D^0$ invariant mass distribution for the sum of $\Bz$ and $\Bp$
candidates in Fig.~\ref{figureAll}e. Due to the proximity of the threshold and to the
fact that the natural width of the $X(3872)$ is comparable to the $\Dbar^{*0} D^{0}$ mass
resolution, there is no easy analytic parametrization of the reconstructed $X(3872)$ mass
spectrum. To measure the mass and width of the $X(3872)$, we generate and reconstruct
high statistics MC samples of $B\to X(3872)K $ events with various masses (from 3872 to
3877~\mevcc) and widths (from 0 to 20~\mev), assuming a pure S-wave decay of a spin 1
resonance. We perform binned extended maximum likelihood fits to the measured $\Dbar^{*0}
D^0$ invariant mass distributions using these different MC samples as signal PDFs
combined with a threshold function for the background. We compare the agreement of each
mass and width hypothesis to the data by computing the $\chi^2$ of the fit for the sum of
bins below 3.9\gevcc. Figure~\ref{chi2} shows the interpolated
$\chi^2-\chi^2_{\mathrm{min}}$ contours versus the simulated masses and widths of the
different signal samples, where $\chi^2_{\mathrm{min}}$ is the $\chi^2$ value for the
best fit. This best fit gives $33 \pm 7$ events in the $X(3872)$ peak, with a statistical
significance of $4.9 \sigma$. Mass and width central values are obtained at the minimum
of the $\chi^2$ distribution, while the errors are given by the extreme points of the
contour in the (mass, width) plane defined at $\chi^2_{\mathrm{min}} + 1$. We obtain a
mass of $(3875.1 {}^{+0.7}_{-0.5} \pm 0.5) \mevcc$ and a width of $(3.0 {}^{+1.9}_{-1.4}
\pm 0.9) \mev$, where the systematic errors include additional contributions from the
choice of the bin width of the invariant mass distribution (0.14 \mevcc\ and 0.07 \mev
respectively on the mass and on the width) and from the fact that in the MC we assume
S-wave \XRes\ decays to $\Dbar^{*0} D^0$ (0.20 \mevcc\ and 0.80 \mev respectively on the
mass and on the width, determined using MC events with $P$-wave \XRes\ decays).
Independently of the mass value, the width measurement is 1.8$\sigma$ away from 0\mev.
The \Bp\ and \Bz\ branching fractions to $\XRes K$ (reported in Table~\ref{BRAll}) are
obtained by fitting the $\Dbar^{*0} D^0$ invariant mass spectra, separately for \Bp\ and
\Bz, choosing the MC sample with $M(X(3872))=3875 \mevcc$ and $\Gamma(X(3872))=3 \mev$,
which is found to give the best fit to the data.

We study the helicity angle of the \XRes, $\theta_{\XRes}$, for the sum of \Bz\ and \Bp\
modes (see Fig.~\ref{figureAll}f). Here, $\theta_{\XRes}$ is defined as the angle between
the $D^0$ or $D^{*0}$ direction and the $B$ direction in the \XRes\ frame. Comparing the
curves obtained with different spin hypotheses with the data distribution, we obtain the
following $\chi^2$/n.d.f.: 9.8/7 for $J^P=1^-$, 3.9/7 for $1^+$ assuming a pure $S$ wave
(flat distribution, not shown in the figure), 2.5/6 for $1^+$ with $S$-$D$ wave
admixture, 5.9/7 for $2^+$ and 2.7/6 for $2^-$. On the basis of this data sample, we
cannot distinguish the different spin assignments. However the $\Dbar^{*0}D^{0}$ decay
would be suppressed by the angular momentum barrier for $J=2$.

The ratio of \XRes\ candidates reconstructed in the $\Dbar^0 D^0 \pi^0$ and $\Dbar^0 D^0
\gamma$ final states is $1.37 \pm 0.56$ (statistical error only), while we expect 1.30
for a decay that proceeds exclusively via a $D^{*0}$ meson. In addition, we measure
parameters which can be used to differentiate various theoretical
interpretations~\cite{X3872Model, X3872Prediction}: $\Delta m$, the mass difference
between the state seen in $B^0$ decays and $B^+$ decays, as well as $R_{0/+}$, the ratio
of branching fractions between $B^0$ decays and $B^+$ decays. Assuming that the signal
seen in $B^0$ decays is not a statistical fluctuation, we obtain $\Delta m$ $=$ $(0.7 \pm
1.9 \pm 0.3) \mevcc$ and $R_{0/+}$ $=$ $1.33 \pm 0.69 \pm 0.43$.

\begin{table*}[htb]
\newcommand\CellTop{\rule{0pt}{2.35ex}}
\newcommand\CellBottom{\rule[-1.2ex]{0pt}{0pt}}
 \begin{center}
 \caption{Fitted event yields, number of events attributed to internal
cross-feed events, efficiencies $\epsilon$ (including intermediate branching fractions),
final branching fractions, \BR, and statistical significances for $B \to D^{(*)}
\DsOneRes$ followed by $\DsOneRes \to D^{*0} K^+$ or $\DsOneRes \to D^{*+} K^0$ (first
part of the table); for $B \to \PsiRes K$ followed by $\PsiRes \to \Dbar^0 D^0$ or
$\PsiRes \to D^- D^+$ (second part of the table); for $B \to \XRes K$ followed by $\XRes
\to \Dbar^{*0} D^0$ (third part of the table). No branching fraction is given in the
absence of signal observation. A 90\% C.L. upper limit on the branching fraction is given
when the statistical significance is less than or equal to $3 \sigma$ (these limits use
the profile likelihood method~\cite{rolke} and include systematic errors). The first
error on \BR\ is statistical and the second is systematic.}
 \label{BRAll} \vskip 0.2cm \begin{tabular}{ll|c|c|c|c|c|c}
 \hline\hline
 \CellTop $B$ mode && Yield & Cross-feed & $\epsilon\ (10^{-4})$
& $\BR\ (10^{-4})$ & Limit $(10^{-4})$ & Significance \\
\hline
\CellTop \modeivDsOne & $[\Dstarz \Kp]$ & 22.4 $\pm$ 6.3 & 0.1 $\pm$ 0.1 & 3.4 & 1.71 $\pm$ 0.48 $\pm$ 0.32 &-& 4.4$\sigma$ \\
\modeviiiDsOne & $[\Dstarz \Kp]$ & 30.9 $\pm$ 7.9 & 1.2 $\pm$ 0.1 & 2.3 & 3.32 $\pm$ 0.88 $\pm$ 0.66 &-& 4.6$\sigma$ \\
\modexviiDsOne & $[\Dstarz \Kp]$ & 29.2 $\pm$ 6.9 & 0.7 $\pm$ 0.1 & 3.5 & 2.16 $\pm$ 0.52 $\pm$ 0.45 &-& 5.6$\sigma$ \\
\modexxDsOne & $[\Dstarz \Kp]$ & 66.7 $\pm$ 13.0 & 6.3 $\pm$ 0.1 & 2.9 & 5.46 $\pm$ 1.17 $\pm$ 1.04 &-& 6.1$\sigma$ \\
\modeviiDsOne & $[\Dstarp K^0]$ & 7.7 $\pm$ 3.1 & 0.0 $\pm$ 0.1 & 0.8 & 2.61 $\pm$ 1.03 $\pm$ 0.31 &-& 3.8$\sigma$ \\
\modexDsOne & $[\Dstarp K^0]$ & 16.8 $\pm$ 5.0 & 0.1 $\pm$ 0.1 & 0.9 & 5.00 $\pm$ 1.51 $\pm$ 0.67 &-& 4.5$\sigma$ \\
\modexviiiDsOne & $[\Dstarp K^0]$ & 7.7 $\pm$ 3.2 & 0.1 $\pm$ 0.1 & 0.9 & 2.30 $\pm$ 0.98 $\pm$ 0.43 &-& 3.3$\sigma$ \\
\modexxiDsOne & $[\Dstarp K^0]$ & 4.8 $\pm$ 2.7 & 0.6 $\pm$ 0.1 & 0.3 & 3.92 $\pm$ 2.46 $\pm$ 0.83 &10.69& 2.3$\sigma$ \\
\hline

\CellTop \modexiPsi & $[\Dzb \Dz]$ & 48.6 $\pm$ 10.2 &-& 9.0 & 1.41 $\pm$ 0.30  $\pm$ 0.22 & - & 6.0$\sigma$ \\
\modeiiPsi & $[\Dzb \Dz]$ & 0.0 $\pm$ 1.5 &-& 1.9 & - & 1.23 & 0.0$\sigma$ \\
\modexiiiPsi & $[\Dm \Dp]$ & 8.9 $\pm$ 3.4 &-& 2.8 & 0.84 $\pm$ 0.32  $\pm$ 0.21 & 1.80 & 3.0$\sigma$ \\
\modeiiiPsi & $[\Dm \Dp]$ & 2.2 $\pm$ 1.6 &-& 0.7 &  0.81 $\pm$ 0.59 $\pm$ 0.09  & 1.88 & 1.6$\sigma$ \\
\hline

\CellTop \modeviX &  [$\Dbar^{*0}D^0$] & 5.8 $\pm$ 2.7 &-& 0.7 & 2.22 $\pm$ 1.05 $\pm$ 0.42 & 4.37& 1.3$\sigma$ \\
\modexxxiX & [$\Dbar^{*0}D^0$] & 27.4 $\pm$ 5.9 &-& 4.3 & 1.67 $\pm$ 0.36 $\pm$ 0.47 & -& 4.6$\sigma$ \\
\hline\hline
 \end{tabular}
 \end{center}
 \end{table*}

In summary, we report the observation of the eight $B$ $\to$ $\Dbar^{(*)} \DsOneRes$
decays, with a \DsOneRes\ mass of $(2534.78 \pm 0.31 \pm 0.40) \mevcc$. We observe the
\PsiRes\ resonance in $B \to \Dbar D K$ decays and measure its mass to be $(3775.5 \pm
2.4 \pm 0.5) \mevcc$. We show that an enhancement of data is observed near the limit of
phase space for the $\Dbar^{*0} D^0$ invariant mass, at a mass of $(3875.1
{}^{+0.7}_{-0.5} \pm 0.5) \mevcc$, with a width of $(3.0 {}^{+1.9}_{-1.4} \pm 0.9) \mev$.
This enhancement could be interpreted as the \XRes, although the observed mass is
$4.5\sigma$ away from the mass measured in the $J/\psi \pi^+ \pi^-$ decay mode. Our mass
value is in good agreement with the value measured by Belle in the $\Dbar^0 D^0 \pi^0$
final state.

We are grateful for the excellent luminosity and machine conditions
provided by our \pep2\ colleagues, 
and for the substantial dedicated effort from
the computing organizations that support \babar.
The collaborating institutions wish to thank 
SLAC for its support and kind hospitality. 
This work is supported by
DOE
and NSF (USA),
NSERC (Canada),
CEA and
CNRS-IN2P3
(France),
BMBF and DFG
(Germany),
INFN (Italy),
FOM (The Netherlands),
NFR (Norway),
MIST (Russia),
MEC (Spain), and
STFC (United Kingdom). 
Individuals have received support from the
Marie Curie EIF (European Union) and
the A.~P.~Sloan Foundation.



\begin{thebibliography}{99}

\bibitem{pdg}
{W.-M. Yao \emph{et al.} (Particle Data Group), J. Phys. G {\bf 33}, 1 (2006).}

\bibitem{patrick}
{B.~Aubert \emph{et al.} (\babar\ Coll.), \jprd{68}, 092001 (2003).}

\bibitem{chunhui}
{B.~Aubert \emph{et al.} (\babar\ Coll.), Phys.\ Rev.\ D, RC~{\bf 74}, 091101 (2006).}

\bibitem{psiBelle}
{K. Abe \emph{et al.} (Belle Coll.), \jprl{93}, 051803 (2004).}

\bibitem{X3872Discovery}
{S.K. Choi \emph{et al.} (Belle Coll.), Phys. Rev. Lett. {\bf 91} 262001 (2003);
B.~Aubert \emph{et al.} (\babar\ Coll.), Phys. Rev. D {\bf 71} 071103 (2005); D. Acosta
\emph{et al.} (CDF II Coll.), Phys. Rev. Lett. {\bf 93} 072001 (2004); V.M. Abazov
\emph{et al.} (D0 Coll.), Phys. Rev. Lett. {\bf 93} 162002 (2004).}

\bibitem{XCDF}
{A. Abulencia \emph{et al.} (CDF Coll.), Phys. Rev. Lett. {\bf 96}, 102002 (2006).}

\bibitem{belleXDDK}
{G. Gokhroo \emph{et al.} (Belle Coll.), \jprl{97}, 162002 (2006). }

\bibitem{X3872Model}
{F.E. Close and P.R. Page, Phys. Lett. {\bf B578} 119 (2004); N.A. Tornqvist, Phys. Lett.
{\bf B590} 209 (2004); L. Maiani \emph{et al.}, Phys. Rev. {\bf D71}, 014028 (2005); Bing
An Li, Phys. Lett. {\bf B605}, 306 (2005); K.K. Seth, Phys. Lett. {\bf B612}, 1 (2005). }

\bibitem{babar}
{B. Aubert \emph{et al.} (\babar\ Coll.), Nucl. Inst. Meth. {\bf A479}, 1 (2002). }

\bibitem{geant4} S.~Agostinelli {\it et al.} (GEANT4 Coll.), Nucl. Instrum. Methods Phys  Res. Sect. A {\bf 506}, 250 (2003).

\bibitem{cleoD0} {
C. Cawlfield {\it et al.} (CLEO Coll.), \jprl{98}, 092002 (2007). }

\bibitem{fox} {
G.C.~Fox and S.~Wolfram, \npb{149}, 413 (1979). }

\bibitem{bes}
{M. Ablikim \emph{et al.} (BES Coll.), Phys. Rev. Lett. {\bf 97} 121801 (2006).}

\bibitem{X3872Prediction} {
E. Braaten and M. Kusunoki, Phys. Rev. {\bf D71}, 074005 (2005); E. Swanson, Phys. Rep.
{\bf 429}, 243-305 (2006).}

\bibitem{rolke}
{W. Rolke, A. Lopez and J. Conrad, Nucl. Instrum. Meth. {\bf A551}, 493 (2005).}

\end{thebibliography}
\end{document}